\documentclass[pra,aps,a4paper,showpacs,floatfix,twocolumn]{revtex4-1}
\usepackage{amsmath}
\usepackage{amsfonts}
\usepackage{amssymb}
\usepackage{graphicx}
\usepackage{dcolumn}
\usepackage{bm}
\begin{document}

\def\Zz{\mathbb{Z}}
\def\T{\hat{T}}

\def\beq{\begin{equation}}
\def\eeq{\end{equation}}
\def\nn{\nonumber}
\def\om{\omega}
\def\a{\alpha}
\def\b{\beta}
\def\g{\gamma}
\def\th{\theta}
\def\eps{\epsilon}
\def\veps{\varepsilon}
\def\l{\lambda}

\def\s{\sigma}
\def\D{\Delta}
\def\d{\delta}

\def\S{\hat{S}}
\def\ph{\phantom}
\def\ra{\rightarrow}

\title{Optimal multi-configuration approximation of an $N$-fermion wave function}
\author{J.~M.~Zhang and Marcus Kollar}
\affiliation{Theoretical Physics III, Center for Electronic Correlations and Magnetism, University of Augsburg, 86135 Augsburg, Germany}

\begin{abstract}
We propose a simple iterative algorithm to construct the optimal multi-configuration approximation of an $N$-fermion wave function. That is, $M\geq N $ single-particle orbitals are sought iteratively so that the projection of the given wave function in the $C_M^N$-dimensional configuration subspace is maximized. The algorithm has a monotonic convergence property and can be easily parallelized. The significance of the algorithm on the study of entanglement in a multi-fermion system and its implication on the multi-configuration time-dependent Hartree-Fock (MCTDHF) are discussed. The ground state and real-time dynamics of spinless fermions with nearest-neighbor interactions are studied using this algorithm, discussing several subtleties.  
\end{abstract}

\maketitle

\section{Introduction}

The wave function of a system of identical fermions must be completely antisymmetric upon exchange of particles. This is a rule of quantum statistics and is independent of the Hamiltonian. The simplest kind of  wave function satisfying this rule is the Slater determinant \cite{slater}: For an $N$-fermion system, take $N$ orthogonal (or at least, linearly independent) orbitals and put one particle in each orbital. After antisymmetrization, the total wave function can be written in the compact form of the determinant of a $N\times N$ matrix, formed by the $N$ orbitals. By construction, a Slater determinant state is a tensor product state conforming to the antisymmetry rule. It is therefore legitimate to consider them as non-correlated or least-correlated states \cite{noquan} for a fermionic system.  

Slater determinant states are simple and serve as the building blocks (basis states) of the multi-fermion Hilbert space. But in a generic system with particle-particle interaction, the wave function (e.g., the ground state) is generally not a Slater determinant. That is, given the wave function, one cannot put it in the form of a Slater determinant, no matter which single-particle orbitals one choose. 

An immediate question is then, whether for a given wave function (say, solved exactly by numerics), one can construct a Slater determinant closest to it in a certain sense. A natural measure of distance or proximity, without reference to any physical quantity, is the overlap (inner product) between the given wave function and the Slater determinant. If the overlap is close to unity, it is plausible to say that the particles are only weakly correlated, though it does not necessarily mean the system is strongly correlated if the overlap is small. 

This question can be generalized. Presumably, in certain systems, 
the fermions may be so strongly correlated that a single Slater determinant is not sufficient to approximate the wave function well. This means, one needs more orbitals than particles to have a bigger, multi-configuration subspace to approximate the given wave function. The question is then, how should we seek the $M> N $ orbitals, or more precisely, the $M$-dimensional single-particle subspace, since the multi-body Hilbert space is determined by the subspace as a whole not by the individual orbitals? 

Several aspects of this problem are significant. First, like the celebrated ``$N$-representability'' problem \cite{yukalov, ando, coleman}, it touches on the fundamental problem of the fine structures of a multi-fermion wave function. Second, it may provide a way to quantify the entanglement in a multi-fermion wave function, which is a topic currently under intensive study \cite{alex, alex2, buchleitner, vollhardt, eriksson}. Third, it helps to better understand the algorithm of multi-configuration time-dependent Hartree-Fock (MCTDHF) \cite{zanghellini,caillat, kato, nest, alon, sakmann}. In this algorithm, the innovative idea is that the single-particle orbitals are not fixed but chosen adaptively, so that the number of orbitals needed to faithfully approximate a many-fermion wave function can be significantly reduced. In practice, the time-evolving wave function is always approximated as a superposition of $C_M^N\equiv  M!/[N!(M-N)!]$ Slater determinants constructed out of $M$ time-dependent orbitals. However the question remains how well the exact wave function can be approximated by using only $M$ orbitals.

In this paper, we propose a simple iterative algorithm as a numerical approach to this problem \cite{uppsala}. The basic idea is to fix all orbitals except for one and optimize it with respect to all other orbitals, and then do the same for the next orbital, and so on. In this process, the overlap increases monotonically, and since it is bounded from above by unity, it is certain to converge. 
The algorithm can be easily understood, implemented, and parallelized. A drawback is that there may exist local maxima. However, at least for the model studied below, there are not too many of them and by running the code several times with different initial values of the orbitals, the global maximum is likely to be found. 

In the following, we first present the algorithm in Sec.~\ref{algorithm}, and then apply it to one-dimensional spinless fermions with nearest-neighbor interaction in Sec.~\ref{model}. We study the convergence of the
algorithm and the Slater approximation of the ground state as well as time-evolved states.  Section \ref{conclusion} contains a discussion of the advantages and limitations of the algorithm.

\section{Iterative algorithm}\label{algorithm}

We start with an $N$-fermion wave function $f(x_1,x_2, \ldots, x_N)$. Here $x_i$ denotes all the degrees of freedom (e.g., position and spin) of the $i$-th particle. In practice, the model is often discretized , so that $x_i$ takes $d$ different values. The $N$-fermion Hilbert space is then of dimension $\mathcal{D} = C_d^N $.
The wave function must satisfy the anti-symmetry condition, i.e., 
\begin{eqnarray}\label{anti}
f(\ldots, x_p , \ldots ,x_q ,\ldots)=-f(\ldots, x_q, \ldots, x_p ,\ldots) ,
\end{eqnarray}
for an arbitrary pair $1\leq p<q  \leq N$.    
Our aim is to find $M\geq N $ orthonormal single-particle orbitals $\{ \phi_n (x) | 1\leq n \leq M \}$, so that the projection of the wave function $f$ in the $C_M^N$-dimensional $N$-fermion subspace can be maximized (the maximum does exist \cite{alexnote}). The latter is spanned by the Slater determinants 
\begin{eqnarray}
S_J (x_1, \ldots, x_N) = \frac{1}{\sqrt{N!}} \det \{\phi_{j_p}(x_q) \}.
\end{eqnarray}
Here $J$ denotes an ordered $N$-tuple $(1\leq j_1<j_2 \ldots < j_N \leq M )$ and the element of the matrix in the $p$th row and $q$th column is $ \phi_{j_p}(x_q) $. An arbitrary normalized wave function in this subspace is of the form 
$ W = \sum_J  C_J S_J $, 
with $\sum_J |C_J|^2 =1$. The overlap between $W$ and the target state $f$, the absolute value of which is to be maximized, is $  \langle f|W \rangle = \sum_J C_J \eta_J  $. Here 
\begin{eqnarray}\label{etaj}
\eta_J &\equiv & \langle f | S_J \rangle =\sqrt{N!} \sum_{x_1}\sum_{x_2}\ldots \sum_{x_N} f^* (x_1, x_2, \ldots, x_N )  \nonumber \\
& & \quad \quad \quad \quad \times \phi_{j_1}(x_1) \phi_{j_2} (x_2) \ldots \phi_{j_N} (x_N),
\end{eqnarray}
by using  the antisymmetry (\ref{anti}) of $f$. 
Simple as it is, Eq.~(\ref{etaj}) is crucial for the algorithm below. The point is that on the right hand side, each orbital appears at most once. 
By the Cauchy-Schwarz inequality, we have 
\begin{eqnarray}
|\langle f|W \rangle |^2\leq  (\sum_J |C_J|^2 ) (\sum_J |\eta_J|^2 )= \sum_J |\eta_J|^2 \equiv \mathcal{I}.
\end{eqnarray}
The equality can be achieved if we choose $C_J \propto \eta_J^* $. For a given set of orbitals $\{ \phi_n |1\leq n \leq M \}$, the maximal value of $|\langle f|W \rangle |^2 $ is then $\mathcal{I} $. The problem is thus one of choosing the orbitals to maximize  $\mathcal{I} $.

For this purpose, let us first fix the orbitals $\{ \phi_2, \ldots, \phi_M\}$, and try to find an optimal $\phi_1$. We have 
\begin{eqnarray}\label{twopart}
\mathcal{I} &=& \sum_{ J|_{j_1\geq 2}} |\eta_J|^2+  \sum_{ J|_{j_1=1} } |\eta_J|^2 \nonumber \\
&=& \sum_{J|_{j_1\geq 2}} |\eta_J|^2+  \sum_{J|_{j_1=1}} |\langle g_J | \phi_1\rangle |^2,
\end{eqnarray}
where the single-particle function $g_J$ is defined as 
\begin{eqnarray}\label{gfun}
g_J (x) &\equiv & \sqrt{N!} \sum_{x_2} \ldots \sum_{ x_N}  f (x, x_2, \ldots, x_N ) \nonumber \\
& &  \quad \quad \quad \quad \times \phi_{j_2}^* (x_2) \ldots \phi_{j_N}^* (x_N).
\end{eqnarray}
Since the first term in (\ref{twopart}) does not depend on $\phi_1$ while the second term does, the task is to maximize the second term in (\ref{twopart}) under the constraint of $\phi_1 \perp  \langle  \phi_2,\phi_3, \ldots, \phi_M \rangle $.
Let $P = \sum_{2\leq n \leq M} |\phi_n\rangle \langle \phi_n | $  be the projection operator onto the subspace spanned by the orbitals $\{ \phi_2, \ldots, \phi_M\}$. We have then 
\begin{eqnarray}\label{quad}
\mathcal{I} &=&  \sum_{J|_{ j_1\geq 2 }} |\eta_J|^2+  \sum_{ J|_{j_1 =1}} \langle \phi_1 | (1-P)|g_J \rangle \langle g_J |(1-P) |\phi_1 \rangle \nonumber \\
&\equiv & \sum_{ J|_{j_1\geq 2} } |\eta_J|^2+   \langle \phi_1  | \hat{T}_1  |\phi_1 \rangle .
\end{eqnarray} 
Here $\hat{T}_1 \equiv \sum_{J|_{j_1 =1}}  |\tilde{g}_J \rangle \langle \tilde{g}_J  | $ and $|\tilde{g}_J\rangle = (1-P) |g_J\rangle $. The eigenvector of $\hat{T}_1$ corresponding to the largest eigenvalue is the optimal $\phi_1$ with respect to the given orbitals $\phi_2$, $\phi_3$, $\ldots$, $\phi_M$. 

By optimizing $\phi_1$, $\mathcal{I} $ has been increased. Now the point is that, while $\phi_1$ is optimized with respect to $\phi_2$, $\ldots$, $\phi_M$, orbital $\phi_M$ is \textit{not} optimized with respect to $\phi_1$, $\ldots$, $\phi_{M-1}$. Therefore, we can turn to $\phi_M$ and optimize it and this would increase $\mathcal{I} $ again. In practice, we can make a circular shift $\phi_i \rightarrow \phi_{i+1}$ after optimizing $\phi_1$ and optimize $\phi_1$ again.
In this way, the value of $\mathcal{I} $ increases monotonically and it will definitely converge since it is up bounded by unity. 

Technically, the most time-consuming part of the algorithm is 
calculating the $g$'s in (\ref{gfun}). In practice, especially when the wave function is represented in second quantization, the value of $f$ is directly available only at points with $x_1 < x_2 \ldots < x_N$. By using the anti-symmetry property of the function, the summation in (\ref{gfun}) can be converted into a form using only these points,
\begin{eqnarray}
g_J(x) &=& \sqrt{N!} \sum_{x_1 < x_2\ldots < x_N } f(x_1 ,x_2, \ldots, x_N) \nonumber \\
& & \quad \quad \quad \quad \;\times \sum_{m=1}^N (-)^{m-1} \delta_{x= x_m } \det \{Y_m \}.
\end{eqnarray}
Here $Y_m$ is the $(N-1)\times (N-1)$ matrix formed by taking  all the rows numbered $\{x_1, x_2, \ldots, x_N \}$ except for the $x_m$-th row of the $d\times (N-1)$ matrix  $[\phi_{j_2}, \phi_{j_3},\ldots, \phi_{j_N}]$. Since the complexity of calculating the determinant is $O(N^3)$, the complexity of calculating one $g_J$ is $O(\mathcal{D} N^4)$. The scaling is not so good, but the summation over $x_1 < x_2\ldots < x_N$ can be easily parallelized. 

\subsection{Single-configuration case ($M=N$)}

The algorithm above simplifies in the special case of $M=N$. In this case, there is only one $J$, i.e., $J=(1,2,\ldots, N)$, and it is readily seen that $g_J$ is orthogonal to $\{ \phi_2, \ldots, \phi_N \}$ \textit{automatically} due to the antisymmetry  of $f$. Therefore, in this case, the optimal $\phi_1$ is simply proportional to $g_J$. Furthermore, if $f$ is a Slater determinant constructed out of orbitals $\{ \psi_1 , \ldots, \psi_N \}$, the algorithm will return $\mathcal{I}=1$ in at most $N$ steps. The reason is that $g_J$ belongs to the subspace spanned by the orbitals $\{ \psi_1 , \ldots, \psi_N \}$, and therefore, after $N$ steps, all the $\phi$'s belong to this subspace. Since they are orthogonal to each other, the $\phi$'s span the same subspace as the $\psi$'s and construct the same Slater determinant $f$.

The optimal orbitals for $N=M$ are called Brueckner orbitals \cite{brueckner1, brueckner2}. They have some useful properties and hence have long been sought-after \cite{nesbet}. In Ref.~\cite{smith} an algorithm similar to ours was proposed, with the difference that the $N$ orbitals are updated all at the same time, while in our case they are updated one by one, yielding monotonous improvement. 

\subsection{The two-fermion or two-boson case ($N=2$)}\label{caseofN2}

For the two-fermion case ($N=2$), the problem can be solved in closed form. First, note that a two-fermion wave function has the canonical form \cite{rmp57,eberly,lowdin,coleman}
\begin{equation}\label{canonicalf}
f(x_1, x_2) = \sum_\alpha \sqrt{ C_{\alpha }} \psi_{2\alpha -1 } \wedge \psi_{2\alpha }  ,
\end{equation}
where $  \psi_{2\alpha -1 } \wedge \psi_{2\alpha } $ denotes the Slater determinant constructed out of the two orbitals $\psi_{2\alpha-1}$ and $\psi_{2\alpha }$, i.e.,
\begin{eqnarray}
& & \psi_{2\alpha -1 } \wedge \psi_{2\alpha }|_{(x_1, x_2)}  \nonumber \\
&\equiv & \frac{1}{\sqrt{2}} \left[ \psi_{2\alpha -1 }(x_1)  \psi_{2\alpha }(x_2)- \psi_{2\alpha -1 }(x_2)  \psi_{2\alpha }(x_1)  \right].\quad 
\end{eqnarray}
The orbitals $\psi_\alpha $ are orthonormal. The coefficients $C_\alpha$ are positive, ordered in descending order, and satisfy the normalization condition $\sum_\alpha C_\alpha =1$. The physical meaning of the $\psi$'s and the $C$'s are apparent by the expression of the one-particle reduced density matrix $\rho_1\equiv 2f^T f^*$ (here $f$ understood as a $d\times d $ matrix),
\begin{eqnarray}\label{rhof}
\rho_1(x,y) = \sum_\alpha {C_\alpha} \left[\psi_{2\alpha-1}(x) \psi_{2\alpha-1}^*(y)+ \psi_{2\alpha}(x) \psi_{2\alpha}^*(y) \right]. \nonumber 
\end{eqnarray}
That is, $\psi_\alpha$ are the natural orbitals and $C_\alpha$ is the (degenerate) occupation number of the orbitals $\psi_{2\alpha-1}$ and $\psi_{2\alpha}$. 

An approximate wave function can also be reduced to a canonical form and therefore, for $N=2$, taking $M$ odd is wasteful because the same state can be constructed using $M-1$ orbitals. We thus assume $M$ even here. Noting that the projection of an antisymmetric function onto the $C_M^2$-dimensional antisymmetric subspace is the same as its projection onto the whole $M^2$-dimensional tensor product space, 
we see that in the case of $N=2$, $\mathcal{I} $ has the compact form  of
\begin{eqnarray}\label{direct}
\mathcal{I} &=& \sum_{1\leq m,n \leq M} \left|\langle \phi_m \phi_n | f \rangle \right|^2 = tr(A^\dagger A ).
\end{eqnarray}  
Here the $M\times M$ matrix $A$ is defined as 
$A = V^T f^* V$, with $V$ the $d\times M $ matrix whose columns are the $M$ orbitals. We have 
\begin{eqnarray}\label{fan}
\mathcal{I} = tr (V^\dagger f^T V^* V^T f^* V ) \leq   tr (V^\dagger f^T  f^* V )  \leq  \sum_{\alpha =1}^{M/2} C_\alpha.\quad \;\;
\end{eqnarray}
Here the first inequality is due to the fact that $1-V^* V^T$ is a projection operator and thus semi-positive definite, while the second one follows from Ky Fan's inequality \cite{matrix}, which essentially states that the sum of $k$ diagonal elements of a hermitian matrix is upper bounded by the sum of its $k$ largest eigenvalues. Note that both equalities can be achieved if the orbitals are chosen as $\phi_n = \psi_n$, $1\leq n \leq M $. Therefore, for $N=2$, the $M$ orbitals which should be chosen are simply the $M$ most populated natural orbitals.

We note that the analysis above can be adapted to the two-boson case straightforwardly. Given a two-boson wave function $b(x_1, x_2)$, with $b(x_1 , x_2)= b(x_2, x_1)$, the similar question is to find $M\geq 1$ orbitals so that the projection of $b$ in the $C_{M+1}^2$-dimensional permanent space is maximal. To answer this question, first we note that similar to (\ref{canonicalf}), a two-boson wave function also has a canonical form \cite{rmp57, eberly, lowdin},
\begin{eqnarray}
b(x_1, x_2) = \sum_\alpha  \sqrt{D_\alpha} \psi_\alpha (x_1) \psi_\alpha(x_2).
\end{eqnarray}
Here the $\psi$'s are orthonormal and the $D$'s are positive and ordered in descending order and satisfying the normalization condition $\sum_\alpha D_\alpha =1$. 
The one-particle reduced density matrix $\rho_1\equiv 2b^T b^*$ is now 
\begin{eqnarray}\label{rhob}
\rho_1(x,y) = \sum_\alpha 2 {D_\alpha} \psi_{\alpha}(x) \psi_{\alpha}^*(y). \nonumber 
\end{eqnarray}
We again have Eq.~(\ref{direct}) if $f$ is replaced by $b$, and similar to (\ref{fan}), we have 
\begin{eqnarray}\label{fan2}
\mathcal{I}  = tr (V^\dagger b^T V^* V^T b^* V ) \leq   tr (V^\dagger b^T  b^* V )  \leq  \sum_{\alpha =1}^{M} D_\alpha.\quad \;
\end{eqnarray}
The equalities can be taken if the orbitals are taken as the first $M$ natural orbitals, i.e., 
$\phi_n = \psi_n$, $1\leq n \leq M$.

\textit{We have thus shown that for the special case of $N=2$, the orbitals which should be taken are simply the natural orbitals.} This means that more direct methods can be used instead of the iteration algorithm, e.g., full exact diagonalization or the Lanczos method. However, it is nonetheless instructive to see how the algorithm works in the special case of $N=M=2$ (see below), which may hint at the convergence behavior of the algorithm also for the more general case of $N>2$. 

Updating $\phi_1$ with respect to $\phi_2$ means replacing $\phi_1$ by $f \phi_2^*$, up to some normalization factor. Updating $\phi_2$ with respect to the new $\phi_1$ means replacing the old $\phi_2$ by $f f^* \phi_2 =- \frac{1}{2}\rho_1 \phi_2$, up to some normalization factor. Therefore, in the special case of $N=M=2$, the algorithm reduces to the power method for the largest eigenvalue. This method is guaranteed to converge to the right result for a generic initial value of $\phi_2$. But it can be slow if $C_2/C_1$ is close to unity and in this case, it can be replaced by the more efficient Lanczos method \cite{golub}. 

\subsection{An upper bound of $\mathcal{I}_\text{max }$ for $N\geq 3$}
In the proceeding subsection, we have proven that in the two-fermion or two-boson case ($N=2$), the maximal value of $\mathcal{I}$ for a multi-configuration approximation with $M $ orbitals has the following expression,
\begin{eqnarray}
\mathcal{I}_\text{max} = \frac{1}{N} \sum_{i=1}^M \lambda_i ,
\end{eqnarray}
where $\lambda_i $ is the $i$th largest eigenvalue of the one-particle reduced density matrix $\rho_1$. We have $\sum_{i=1}^d \lambda_i = N$.

We now proceed to show that in the more general case of $N\geq 3$, for both fermions and bosons, there holds the inequality 
\begin{eqnarray}\label{myine}
\mathcal{I}_\text{max} \leq  \frac{1}{N} \sum_{i=1}^M \lambda_i .
\end{eqnarray}
To prove this inequality, suppose $\{ \phi_1, \phi_2, \ldots, \phi_M \}$ are a set of (orthonormal) optimal orbitals. We can extend this basis to a full basis $\{ \phi_1, \phi_2, \ldots, \phi_M,\ldots, \phi_d \}$ of the single-particle space. Denote the creation operator corresponding to orbital $\phi_i$ as $a_i^\dagger$. The wave function $f$ can be expanded as 
\begin{subequations}\label{fexp}
\begin{eqnarray}
f & = &  \sum_{J \subseteq A}   C_J |J\rangle +\sum_{J \nsubseteq A }  C_J |J\rangle  ,  \\
 & & |J \rangle \propto a_{j_1}^\dagger a_{j_2}^\dagger \ldots a_{j_N}^\dagger |vac\rangle,\quad \langle J | J \rangle =1.
\end{eqnarray} 
\end{subequations}
Here $A$ is the set $\{ 1, 2, \ldots, M \}$ and $J=\{ j_1, j_2 ,\ldots, j_N \}$ is an $N$-tuple. For fermions, the elements of $J$ are all different while for bosons, some of them can be identical. The notation $J \subseteq A$ means that each element of $J$ is also an element of $A$, otherwise $J \nsubseteq A$ applies. We have 
\begin{eqnarray}\label{imax}
\mathcal{I}_\text{max}=  \sum_{J \subseteq A} |C_J|^2. 
\end{eqnarray}
Now we note that 
\begin{eqnarray}\label{ine}
\sum_{i=1}^M \lambda_i &\geq & \sum_{i=1}^M \langle \phi_i | \rho_1 | \phi_i \rangle \nonumber =\sum_{i=1}^M \langle f | a_i^\dagger a_i |f \rangle  \\
&=& \sum_{J \subseteq A}  N |C_J|^2 + \sum_{J \nsubseteq A }|C_J|^2 \langle J | \sum_{i=1}^M a_i^\dagger a_i | J \rangle . \nonumber \\ 
&\geq & N \sum_{J \subseteq A} |C_J|^2. 
\end{eqnarray}
The first inequality is again a consequence of Ky Fan's inequality while the second one is obvious. Combining (\ref{imax}) and (\ref{ine}), we get (\ref{myine}).

\begin{figure*}[tb]
\includegraphics[width= 0.9 \textwidth ]{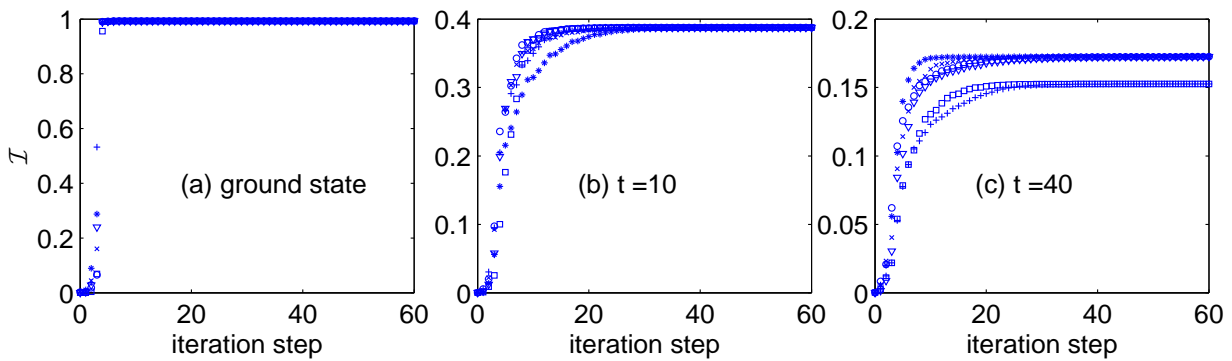}
\caption{(Color online) Convergence behavior of the algorithm  illustrated with three test states which correspond to the three panels one to one. The common parameters are $(N,M,L,U)=(4,4,20,1)$. By one iteration step we mean updating one orbital. In each panel, six different trajectories are shown, which correspond to six different sets of initial orbitals and are differentiated by different markers. The test state in panel (a) is the ground state with $N$ fermions on a $L$-site lattice, while the test states in panels (b) and (c) are the time-evolved state at time $t=10$ and $t=40$, respectively, with the fermions initially confined to the $N$ left most sites of the lattice. See the text in Sec.~\ref{convprop} for more detailed explanation. Note that in panel (c), two trajectories converge to a local maximum. 
\label{conv}}
\end{figure*}

\section{Spinless fermions with nearest-neighbor interaction}\label{model}
We now turn to a specific model in order to check the performance of
  the algorithm and to study physical properties from the perspective
  of entanglement in a multipartite system. We consider spinless (or
  spin polarized) fermions on a one-dimensional $L$-site lattice.
  Specifically, the Hamiltonian is ($\hbar =1$ throughout)
  \begin{eqnarray}\label{H}
    \hat{H }= \sum_{i=1}^{L-1} - (\hat{c}_i^\dagger \hat{c}_{i+1} + \hat{c}_{i+1}^\dagger \hat{c}_i ) + U \hat{n}_i \hat{n}_{i+1}.
  \end{eqnarray}
  Here $\hat{c}_i^\dagger$ ($\hat{c}_i$) is the creation
  (annihilation) operator on site $i $, and $\hat{n}_i\equiv
  \hat{c}_i^\dagger \hat{c}_i$ counts the occupation number. The usual
  anti-commutation relations hold, i.e., $\{ \hat{c}_i, \hat{c}_j \} =
  \{ \hat{c}_i^\dagger, \hat{c}_j^\dagger \}=0$, and $\{ \hat{c}_i,
  \hat{c}_j^\dagger \} = \delta_{ij}$. The first term is the hopping
  term while the second term describes the interaction (with strength
  $U$) between two fermions sitting on adjacent sites. The total
  number of fermions is conserved and will be denoted as $N$. We use
  open boundary conditions.

\subsection{Convergence of the iterative algorithm}\label{convprop}

\begin{figure}[tb]
\includegraphics[width= 0.4 \textwidth ]{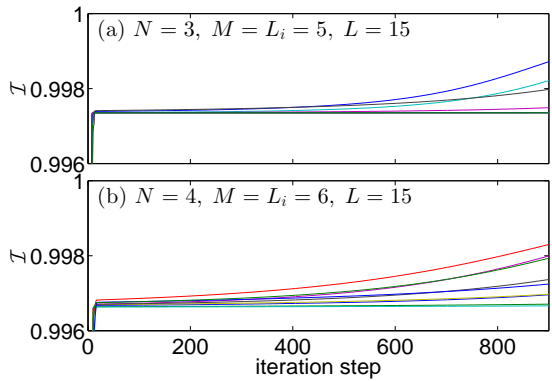}
\caption{(Color online) Convergence of $\mathcal{I}$ can be slow in the late stage. In each panel, the test state $\psi(t=20)$ is from such a scenario. Initially $N$ fermions are confined to the $L_i$ left most sites of a $L$-site lattice and are in the ground state. The interaction strength $U=1$. At $t=0$, the confinement is lifted and the value of $U$ is quenched to zero. In each panel, nine runs with different initial values of the $M=L_i$ orbitals are shown. Rigorously, $\mathcal{I}_\text{max}=1$. Note the scale of the axes. 
\label{slow}}
\end{figure}

  The first step is to check the convergence property of the
  algorithm. Instead of trying the algorithm on artificial states, we
  prefer physical states such as the ground state or a time-evolved
  state after a sudden quench. Specifically, we will use states of the model (\ref{H}).

Three cases are illustrated in Fig.~\ref{conv}. In Fig.~\ref{conv}(a), we have $N=4$ fermions on a lattice with $L=20$ sites and the system is in the ground state. In Figs.~\ref{conv}(b) and \ref{conv}(c), the two test states are from the following scenario. Initially the four fermions are confined to the four left most sites of the 20-site lattice by some wall and, since there is no freedom at all, the system is obviously in a Slater determinant wave function. Then at $t=0$, the wall is suddenly removed and the fermions expand into the whole lattice. Denoting the wave function at time $t$ as $\psi(t)$, the test wave functions in Fig.~\ref{conv}(b) and \ref{conv}(c) are $\psi(t=10)$ and $\psi(t=40)$, respectively. 

For all the three test states, we try to approximate them with a single Slater determinant, i.e., the number of single-particle orbitals is $M=N=4$. Moreover, for each test state, we have run the algorithm with six different initial sets of orbitals $\{ \phi_n| 1\leq n \leq M \}$, marked by different symbols. We see that for all the states and for all the initial values of the orbitals, the value of $\mathcal{I }$ converges to its final value monotonically from below as expected, and moreover, generally in no more than 50 iteration steps, it has converged with good precision. 

Figures \ref{conv}(a) and \ref{conv}(b) demonstrate that the algorithm works well; all six runs converge to the same limiting value.  However, in Fig.~\ref{conv}(c), we have two runs converging to a local maximum. More extensive investigation (with up to hundreds of runs) showed that for the two states in Figs.~\ref{conv}(a) and \ref{conv}(b), the algorithm would never be trapped in a local maximum, and for the state in Fig.~\ref{conv}(c), fortunately, there is only one local maximum and the probability of hitting it is less than $25\%$. 

To deal with the local maxima, we simply run the code multiple times (at least six) \cite{NO} and take the maximal final value of $\mathcal{I}$. Extensive investigation showed that this strategy is a very safe one. At least for the model (\ref{H}), the number of local maxima is small (we have not observed any case with more than one local maxima) and the probability of hitting them is smaller than hitting the global one. Therefore, we are confident that our data are exact. In the following, all data are generated in this way.

Some remarks are in order. In Fig.~\ref{conv}, on the scale of the
  vertical axes, it seems that $\mathcal{I}$ has completely converged
  once the plateau is reached, when in fact a slow drift is still in
  progress. To demonstrate this fact, we have considered the following
  scenario. Suppose initially the $N$ fermions are confined to the
  $L_i$ left most sites and are in the ground state. Then at $t=0$,
  the confinement is lifted and at the same time the value of $U$ is
  quenched to zero.  The subsequent evolution is an evolution of free
  fermions and the wave function can always be exactly represented
  with $L_i$ orbitals, which can be easily solved from the
  single-particle Schr\"odinger equation. That is, for $\psi(t)$ with
  an arbitrary $t$, $\mathcal{I}_{\text{max}} =1$ if $M\geq L_i$.

  We have checked the algorithm against this rigorous result. The
  trajectories of $\mathcal{I}$ are plotted in Fig.~\ref{slow}. There,
  we see that the initial phase of convergence is fast. Regardless of
  the initial values of the orbitals, $\mathcal{I}$ builds up to
  around $0.997$ in 15 iteration steps. However, the subsequent
  process of convergence towards unity is very slow. Hundreds of steps
  are needed to reduce the error by one half even in the best
  case. This late slow improvement is understandable in view of the
  fact that in the $N=M=2$ case, the algorithm reduces to the power
  method, which is well-known to be slow in the rate of convergence.
  Fortunately, the first, relatively fast stage of convergence
  generally can already bring $\mathcal{I}$ sufficiently close to
  $\mathcal{I}_{\text{max}}$, and therefore, the late slow convergence does
  not prevent us from studying physics quantitatively using the
  algorithm.

\subsection{Ground state}
\begin{figure*}[tb]
\includegraphics[width= 0.45 \textwidth ]{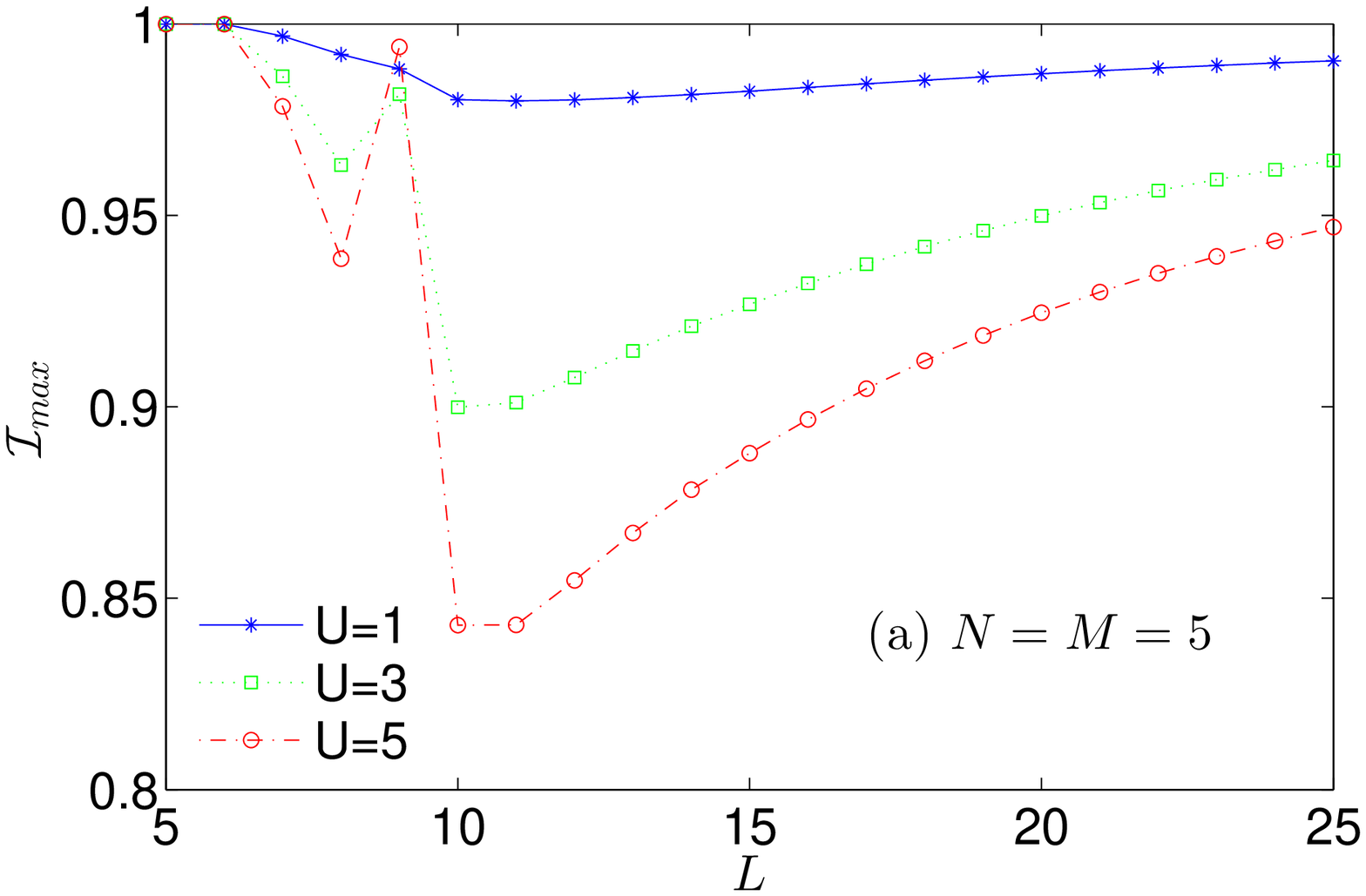}
\includegraphics[width= 0.45 \textwidth ]{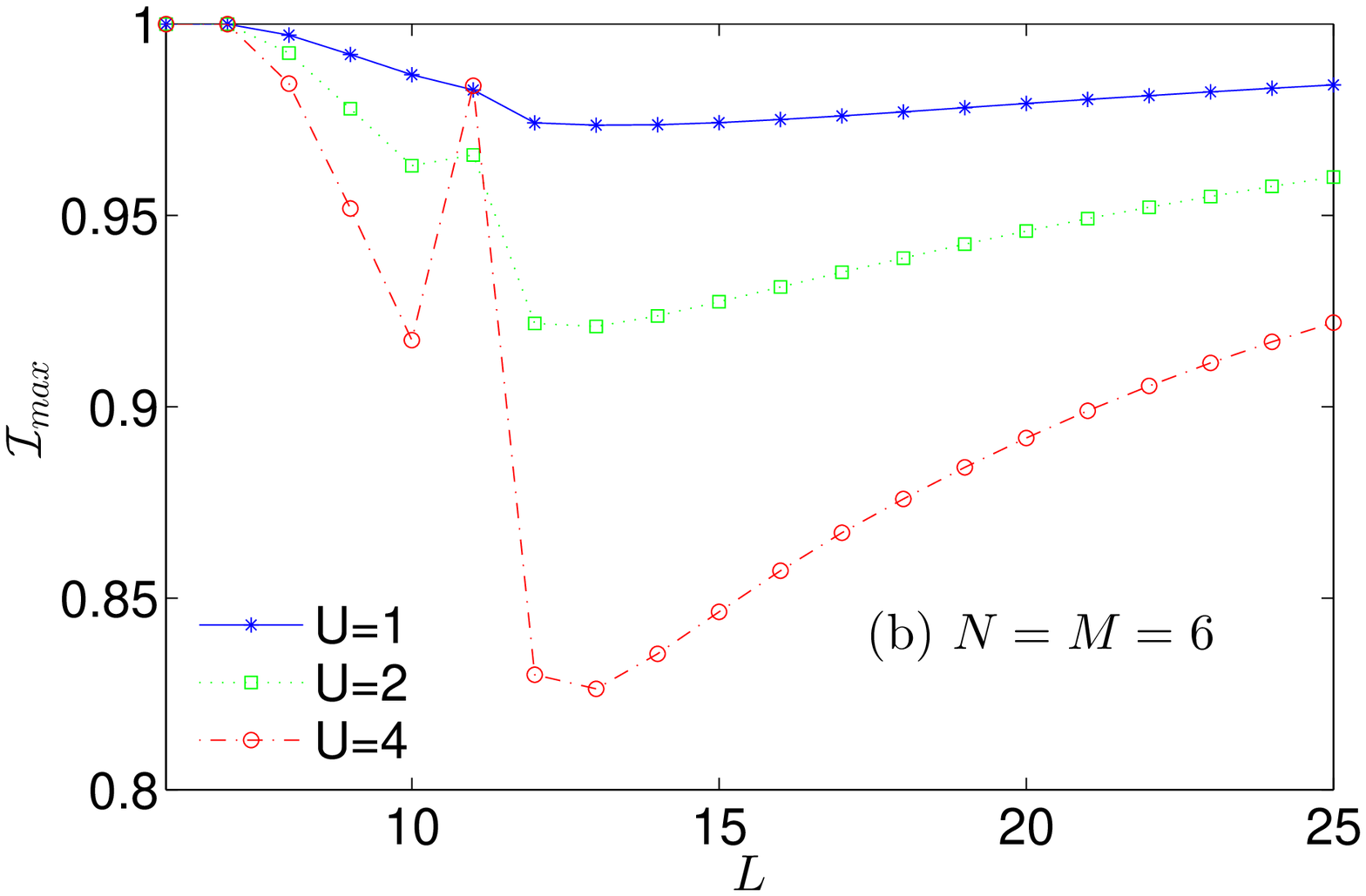}
\caption{(Color online) The maximal value of $\mathcal{I}$ for the ground state of the model (\ref{H}), as a function of the size $L$ of the lattice, in the repulsive interaction case ($U>0$). The number of fermions $N$ is $N=5$ in (a) and $N=6$ in (b), respectively. The number of orbitals is always chosen as $M=N$, i.e., the ground state is approximated with a single Slater determinant. Note that, in both panels, as the repulsive interaction strength $U$ increases, $\mathcal{I}_\text{max}$ develops a local maximum at $L= 2N-1$, as a result of the formation of the charge-density-wave order. Moreover, $\mathcal{I}_\text{max}=1$ rigorously when $L=N$ or $N+1$, as a consequence of a theorem \cite{ando}.
\label{gspositive}}
\end{figure*}

\begin{figure*}[tb]
\includegraphics[width= 0.45 \textwidth ]{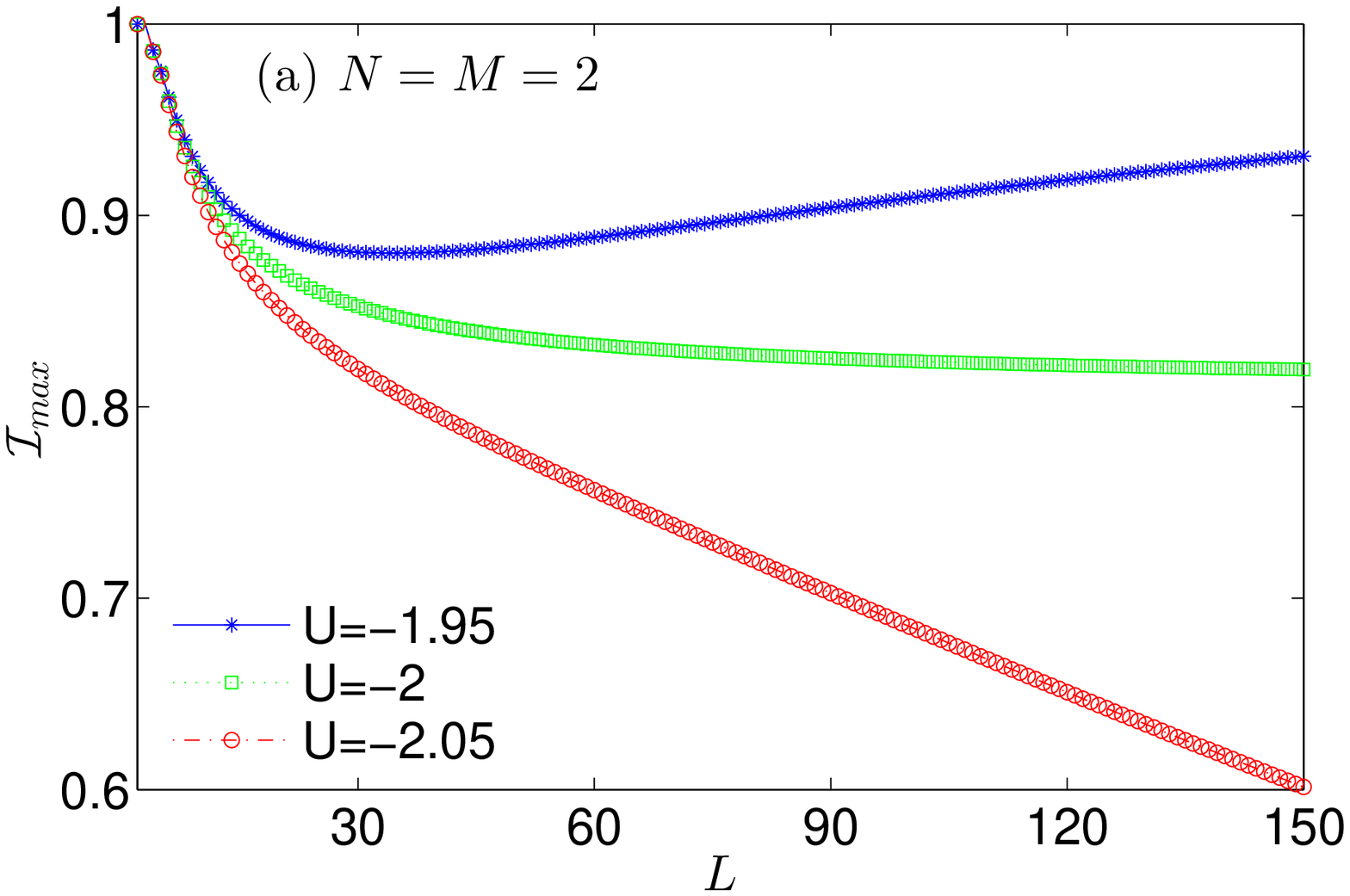}
\includegraphics[width= 0.45 \textwidth ]{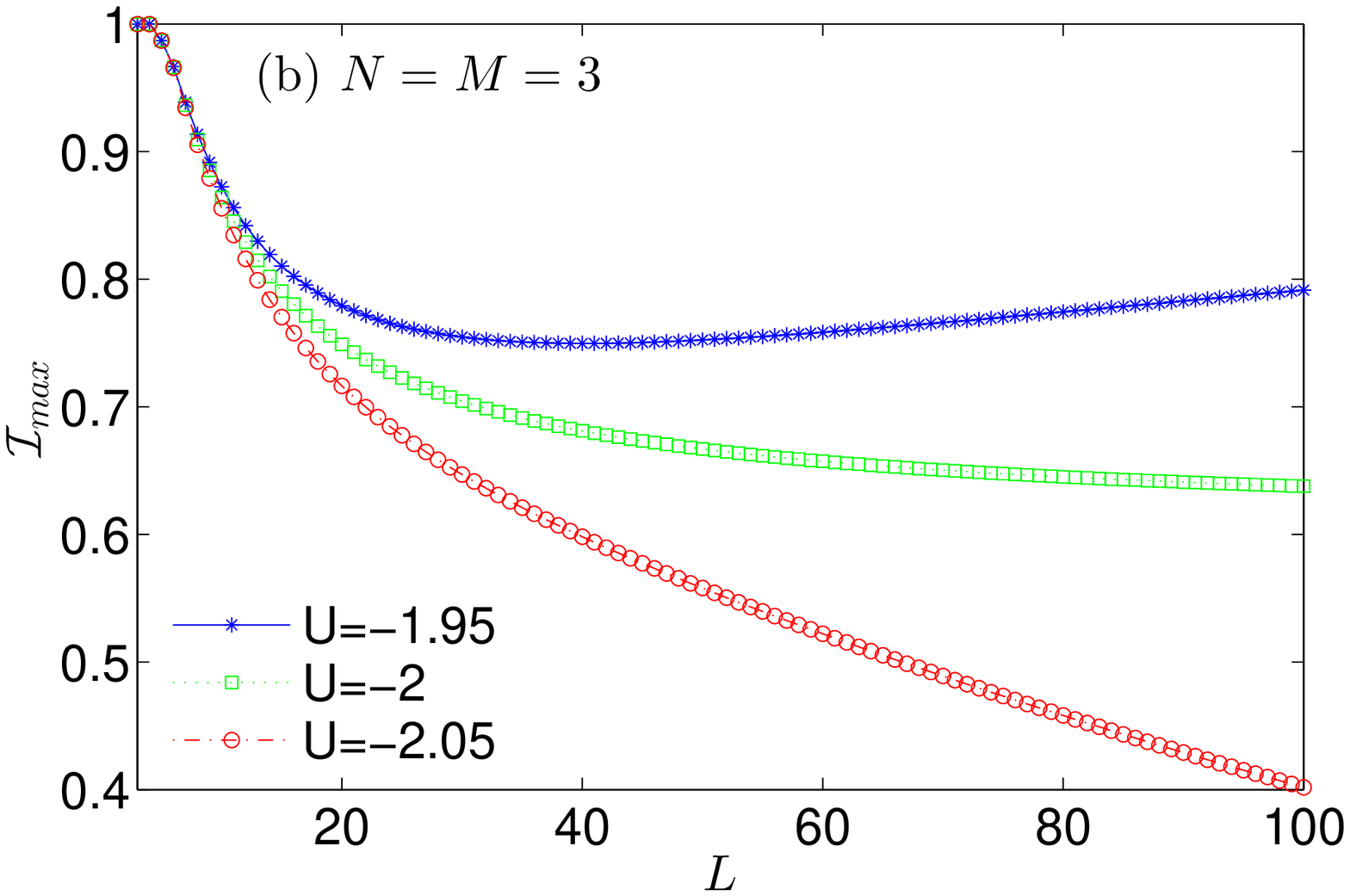}
\caption{(Color online) The maximal value of $\mathcal{I}$ for the ground state of the model (\ref{H}), as a function of the size $L$ of the lattice, in the case of attractive interaction ($U<0$). The number of fermions $N$ is $N=2$ in (a) and $N=3$ in (b), respectively. The number of orbitals is always chosen as $M=N$, i.e., the ground state is approximated with a single Slater determinant. Note the bifurcation at $U_c =-2$ in each panel. For $|U|<|U_c|$, there is no binding of two particles and $\mathcal{I}_\text{max} $ converges to unity as $L\rightarrow +\infty$; while for $|U|> |U_c|$, two particles can bind into a pair and $\mathcal{I}_\text{max} $ converges to zero in this limit. 
\label{gsnegative}}
\end{figure*}

Having established the reliability of the algorithm, we now proceed to study physical properties of the model (\ref{H}) using it. The first object that we consider is the ground state. The question is, in the presence of interaction, to what extent can the ground state be approximated using a Slater determinant. How does the strength of the interaction and the volume accessible to the fermions (i.e., the lattice size $L$) influence it? 
It turns out that the answer depends on the sign of $U$. 

Let us treat the case of positive $U$ first. For positive $U$, two
  limits allow for a simple picture. In the limit of $L=N$, the
  fermions have no freedom to distribute themselves and the wave
  function is exactly in a Slater determinant form. In the opposite
  limit of $L\rightarrow +\infty$, the fermions have little chance to
  encounter each other and interact and the ground state should can be
  well approximated by a Slater determinant. Therefore, in the two
  limits, it is expected that $\mathcal{I}_{\text{max}}=1$, and there should
  be a minimum of $\mathcal{I}_{\text{max}} $ at some intermediate $L$. This
  is indeed what is observed in Fig.~\ref{gspositive}. There, in the
  two panels, which correspond to two different values of $N$ ($N=5$
  and $N=6$), we see that regardless of the value of $U$,
  $\mathcal{I}_{\text{max}} $ tends to unity at the two extremes and develops
  a minimum in between. Note that for $L=N$ and
  $N+1$, $\mathcal{I}_{\text{max}} =1$ exactly. This is trivial for $L=N$.
  For $L=N+1$, it is a consequence of a theorem \cite{ando} (see also
  Appendix~\ref{appA}), i.e., if there are only $N+1 $ orbitals available
  to $N$ fermions, then the wave function must be a Slater
  determinant.  In this paper, we shall refer to this theorem as the
  $N$-in-$(N+1)$ theorem.

  A noticeable fact in Fig.~\ref{gspositive} is that, at $L=2N-1$, as
  $U$ increases, a local maximum develops. As $U\rightarrow +\infty$,
  this local maximum will go to unity. This local maximum is a
  geometric effect. For $L=2N-1$, the $N$ fermions can reside at every
  other lattice site, thus forming a charge-density wave (CDW). In this configuration, they can avoid to
  interact with each other. This unique configuration has the lowest
  interaction energy and in the large $U$ limit, it would be a good
  approximation to the ground state and therefore
  $\mathcal{I}_{\text{max}} \rightarrow 1$. On the contrary, for other
  values of $L$, there are several configurations degenerate in the
  interaction energy. In the large $U$ limit, they hybridize into the
  ground state and therefore the latter has much worse approximation.
  At $L=2N-1$, the picture is that as $U$ increases from zero to
  infinity, the Slater determinant approximation first deteriorates and then becomes perfect again. 

  The negative $U$ case is also interesting. For negative $U$, in the
  ground state, there is the possibility of binding of particles if
  the attraction is strong enough. That is, all or some the particles
  are bound together. Once this happens, the correlation between the
  particles is strong and the Slater approximation will get worse.

  For the two-fermion case, for the model (\ref{H}) on a infinite
  lattice, it is easy to show that the critical value of $U$ for the
  ground state to be a bound fermion-pair is $U_c= -2$.  For $|U|<
  |U_c|$, the attraction is insufficient to bind the two fermions.
  They interact but are delocalized with respect to each other. Like
  in the positive $U$ case, we expect $\mathcal{I}_{\text{max}} $ to go to
  unity in the $L\rightarrow +\infty$ limit.

  However, for $|U|> |U_c|$, the two fermions form a pair and hops on
  the lattice as a whole. In terms of the wave function $f(x_1, x_2)$,
  it has a band profile along the diagonal $x_1= x_2$.  This in turn
  means that the one-particle reduced density matrix $2f^\dagger f$
  has also a band profile and the largest magnitude of its element is
  on the order of $1/L$. The magnitude of the element should decay
  exponentially away from the main diagonal. Therefore, by the
  Gershgorin disk theorem \cite{circle}, the largest eigenvalue of
  $2f^\dagger f$ should be on the order of $1/L$. By the analysis in
  Sec.~\ref{caseofN2}, $\mathcal{I}_{\text{max}} $ should be on the order of  $1/L$. In particular, it decays to zero in the $L \rightarrow
  +\infty$ limit.

  These predictions are verified by the algorithm \cite{scaling}. In Fig.~\ref{gsnegative}(a), with $N=2$ and for three different values
  of $U $, $\mathcal{I}_{\text{max}} $ is plotted as a function of $L$. The
  bifurcation is clear. At the critical value of $U=- 2$,
  $\mathcal{I}_{\text{max}} $ converges to a finite value slowly. On the two sides, it either converges to unity or to zero.
  In Fig.~\ref{gsnegative}(b), the $N=3$ case is
  studied. We again see the bifurcation behavior at $U=-2$. It signals
  that beyond this critical value of $U$, in the ground state two of
  the three fermions are bound together, although not necessarily all
  of them.

\subsection{Real-time dynamics after a quantum quench}

\begin{figure*}[tb]
\includegraphics[width= 0.45 \textwidth ]{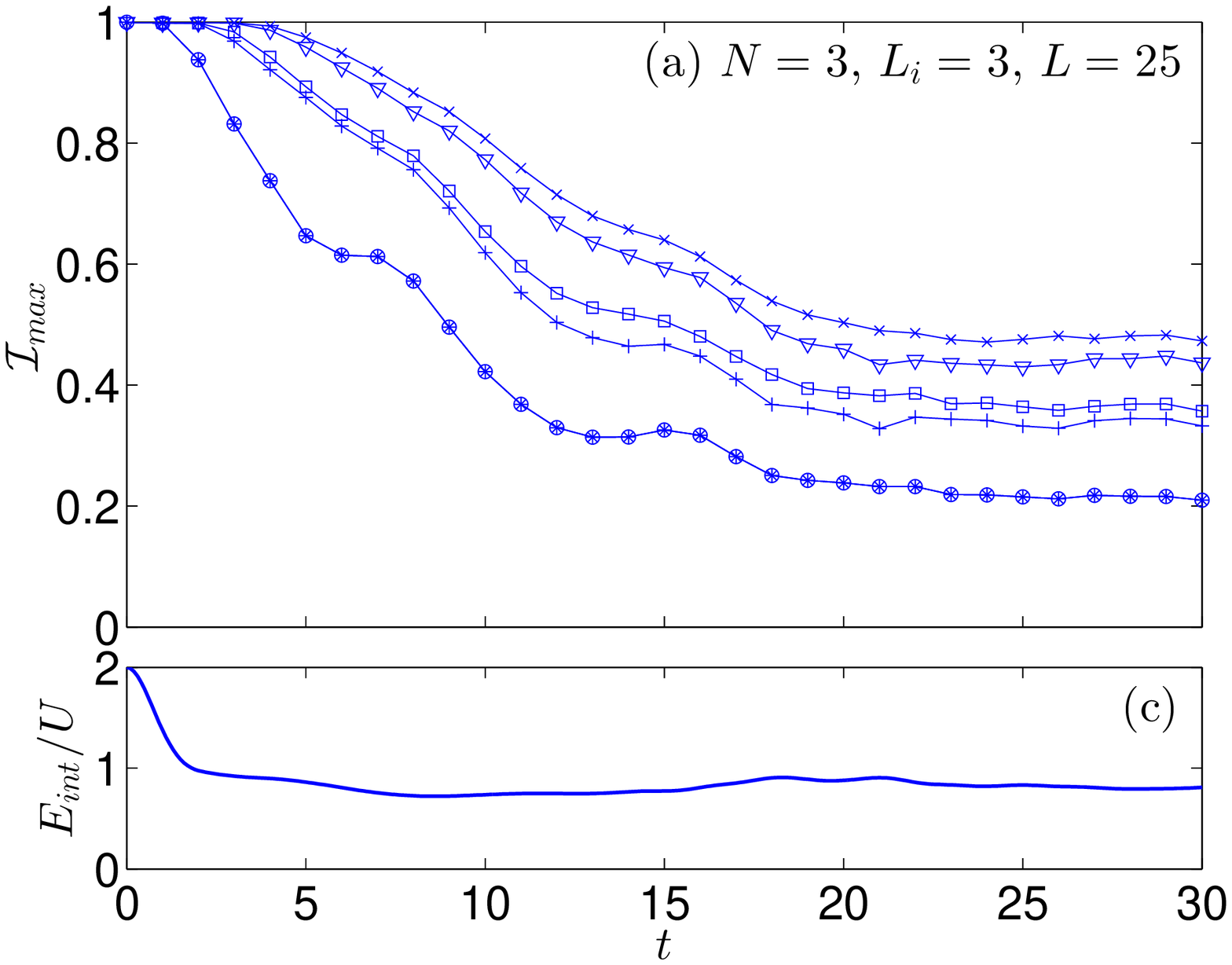}
\includegraphics[width= 0.455 \textwidth ]{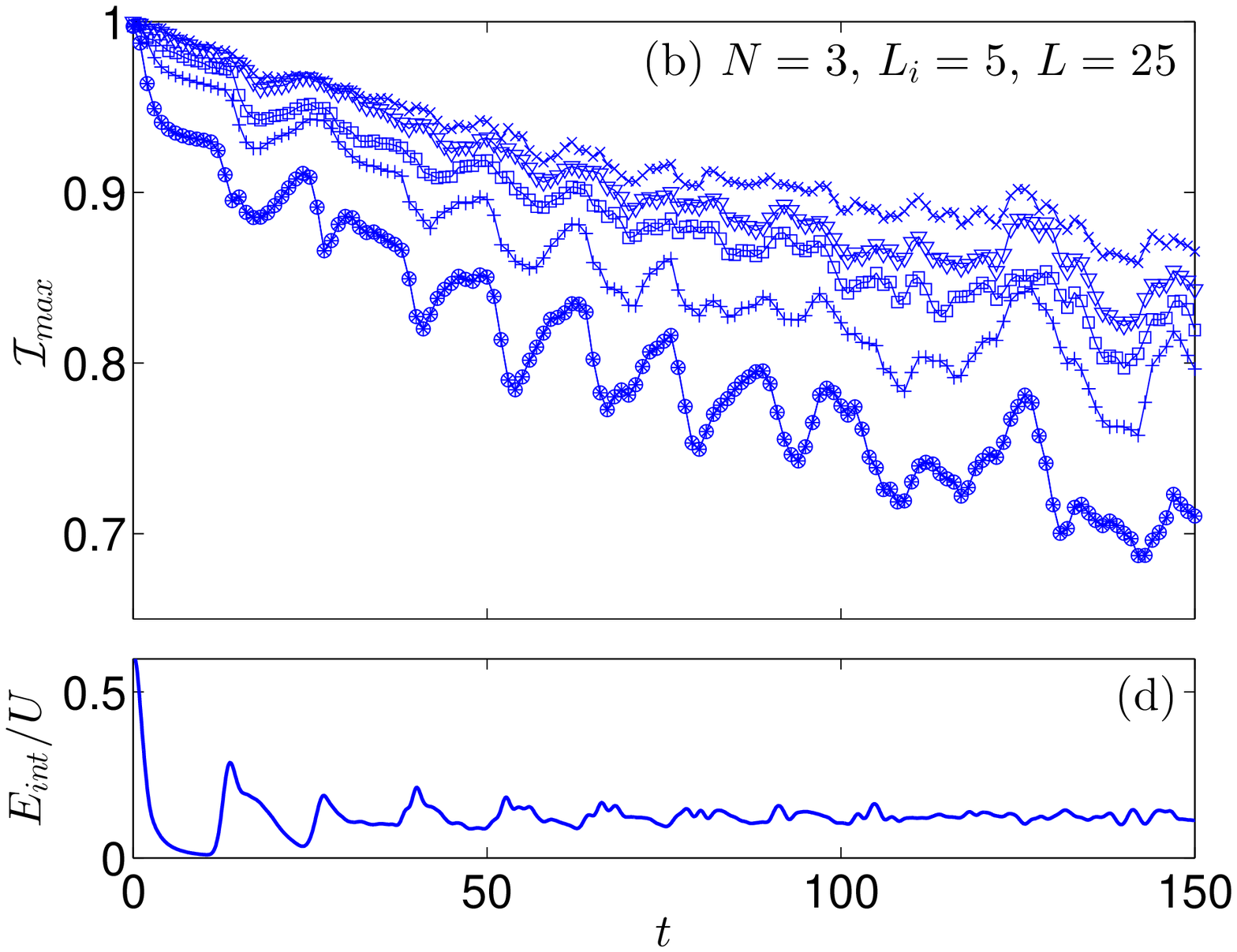}
\caption{(Color online) Evolution of the fidelity $\mathcal{I}_\text{max} $ after a quantum quench. Initially the $N=3$ fermions are confined to the (a) $L_i=3$ or (b) $L_i= 5$ left most sites of a lattice of $L=25$ sites, and are in the ground state.  The interaction strength $U=1$. The confinement is lifted at $t=0$. In each panel, from bottom to up, the six lines correspond to $M$ from 3 to 8. Note that the lowest line is two lines in coincidence, as a result of a theorem \cite{ando, right}. Also note the different scales of the axes in the two panels. Panels (c) and (d) show the time evolution of the interaction energy in panels (a) and (b), respectively. Note the periodic peak-dip correspondence in (b) and (d).
\label{evoeta}}
\end{figure*}

\begin{figure}[tb]
\includegraphics[width= 0.4 \textwidth ]{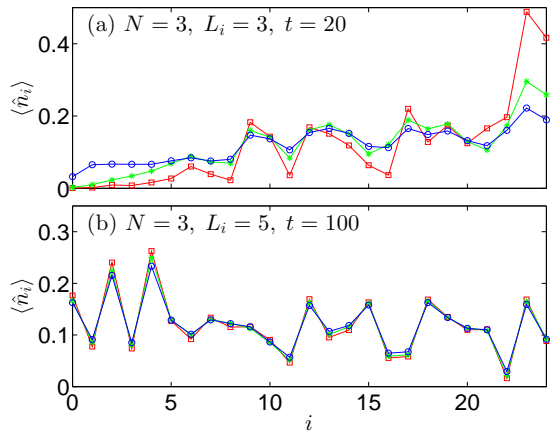}
\caption{(Color online) Density distribution $\langle \hat{n}_i \rangle $ of the exact wave function $\psi(t)$ and its optimal multi-configuration approximation. The two panels correspond to the panels (a) and (b) in Fig.~\ref{evoeta}, respectively. In each panel, the blue line with circles corresponds to the exact wave function $\psi(t)$, the red line with squares corresponds to the approximate wave function with $M=3$ orbitals, while the green line with asterisks to the approximate wave function with $M=8$ orbitals. 
\label{dencomp}}
\end{figure}

Upon time evolution, the particles collide  with each other.
  Therefore, the wave function may become very complicated when
  represented with Slater determinants. An important type of dynamics
  starts from a wave function which is or can be well approximated by
  a Slater determinant. With time, the wave function is likely to become
  entangled and the approximation will deteriorate. The rate and the
  extent of the developing entanglement is interesting in its own
  right, and will be discussed in the context of the MCTDHF algorithm
  and its accuracy.

\subsubsection{Quantum quench dynamics}

Consider the following scenario. Initially the $N$ fermions are
  confined to the $L_i$ left most sites of the lattice, and they are in
  the ground state. Then suddenly at $t=0$, the confinement is lifted
  and the fermions start expanding into the rest of the lattice. The
  wave function at time $t$ can be solved by exact diagonalization
  \cite{ejp}, and then the optimal multi-configuration approximation
  can be searched using the iterative algorithm.

In Fig.~\ref{evoeta}, two typical examples are shown. The two panels share the same number of fermions ($N=3$), the same total number of sites $(L=25)$, and the same value of interaction strength ($U=1$). The difference comes from the  value of $L_i$. In panel (a), $L_i=3$ while in panel (b), $L_i=5$. Therefore, the only difference is in the initial state. For each time slice, the state is approximated using various numbers of orbitals, i.e., with from $M=3$ up to $M=8$ orbitals, which correspond to the lines from bottom to top (but note that the line with $M=3$ coincides with the line with $M=4$ due to the same reason as in Figs.~\ref{gspositive} and \ref{gsnegative}).
 
We see that in case (a), the value of $\mathcal{I}_\text{max} $ decays quickly. This is especially significant for $M=3$ (as well as $M=4$). By increasing $M$, the decay of $\mathcal{I}_\text{max}$ is slowed down but not dramatically. Moreover, the deviation from unity does not reduce significantly as $M$ increases. All these facts indicate that for this initial state, the fermions 
become strongly entangled quickly, in the sense that they can no longer be described using a small number of orbitals.  

Case (b) is quite different. Now the fidelity $\mathcal{I}_\text{max} $ decays much slower and even in the single configuration approximation and even at time $t=100$, $\mathcal{I}_\text{max} $ is close to $0.8 $. That is, in a large time interval, the wave function can always be well approximated using a very limited (even minimal) number  of orbitals. 

It is not obvious why the approximation of the initial state in (a)
  deteriorates quickly but the one in (b) only slowly and weakly.
  However, we note that the interaction clearly plays a role. In Figs.~\ref{evoeta}(c) and \ref{evoeta}(d), the expectation value of the interaction energy in Figs.~\ref{evoeta}(a) and \ref{evoeta}(b), respectively, 
is plotted against time. While $E_{int}/U$ is close to unity all the time in (c), which means the probability of finding a pair of fermions interacting with each other is high, it is smaller by one order of magnitude in (d) most of the time. Of course, without interaction, the value of $\mathcal{I}_\text{max} $ will remain constant. This may explain the relatively slow and weak decay of $\mathcal{I}_\text{max}$ in (b).

A subtle fact in Figs.~\ref{evoeta}(b) and \ref{evoeta}(d) is also noteworthy. In Fig.~\ref{evoeta}(d), we see that the interaction energy $E_{int}$ shows periodic peaks (with a period $\simeq 13$). The reason behind is that the particles bounce back and forward between the lattice ends. When they concentrate at an end [see Fig.~\ref{den}(b3)], the interaction energy develops a peak, and accordingly, the value of $\mathcal{I}_\text{max}$ in the case of $M=3$ or $4$ drops down relatively quickly. Interestingly, sometimes when $E_{int}$ decreases (the particles disperse again), the value of $\mathcal{I}_\text{max}$ moves up again (revival), and we observe a clear peak-dip correspondence for $t\leq 100$. We then see that under time-evolution, $\mathcal{I}_\text{max}$ does not necessarily decrease monotonically.

So far, we have focused on $\mathcal{I}_{\text{max}}$, which appears
  only as a mathematical construct. To understand its relation to
  physical quantities, it is necessary to compare a given wave
  function and its optimal multi-configuration approximation (in the
  sense of inner product) regarding a specific observable. Does the
  value of $\mathcal{I}_{\text{max}}$ have any implication on it?
  Especially, does a $\mathcal{I}_{\text{max}}$ close to unity imply
  that the approximate wave function is a good one as regards a
  \textit{generic} physical quantity too?  Fig.~\ref{dencomp} shows
  the density distribution $\langle \hat{n}_i \rangle $ of the exact
  wave function $\psi(t)$ and those of its optimal multi-configuration
  approximations. In each panel of Fig.~\ref{dencomp}, a wave function
  from the corresponding panel of Fig.~\ref{evoeta} is taken, for
  which two optimal approximations with $M=3$ or $8$ are then
  constructed using the algorithm. In Fig.~\ref{dencomp}(a), we see
  that the $M=3$ approximation differs from the exact distribution
  significantly. The $M=8$ approximation becomes better but the
  discrepancy is still apparent. This is in agreement with the low
  values of $\mathcal{I}_{\text{max}}$ in Fig.~\ref{evoeta}(a). There,
  at $t=20$, $\mathcal{I}_{\text{max}}$ is as small as $0.21$ for
  $M=3$ and only $0.5$ for $M=8 $. 

On the contrary, in Fig.~\ref{dencomp}(b), at $t=100$, we see that the density distributions of both approximations are very close to the exact one. This is quite remarkable. Note that at $t=100$, the interaction energy $E_{int}$ has peaked eight times. Yet a single Slater determinant  suffices to capture the density distribution of the exact wave function quantitatively. This is of course a favorable situation for the MCTDHF algorithm. It is also in agreement with the high values of $\mathcal{I}_\text{max}$ in Fig.~\ref{evoeta}(b). In Appendix \ref{appB}, we  show that closeness in inner product implies closeness in density distribution, though the reverse is obviously not true.

\subsubsection{Consequences for MCTDHF}

\begin{figure*}[tb]
\includegraphics[width= 0.45 \textwidth ]{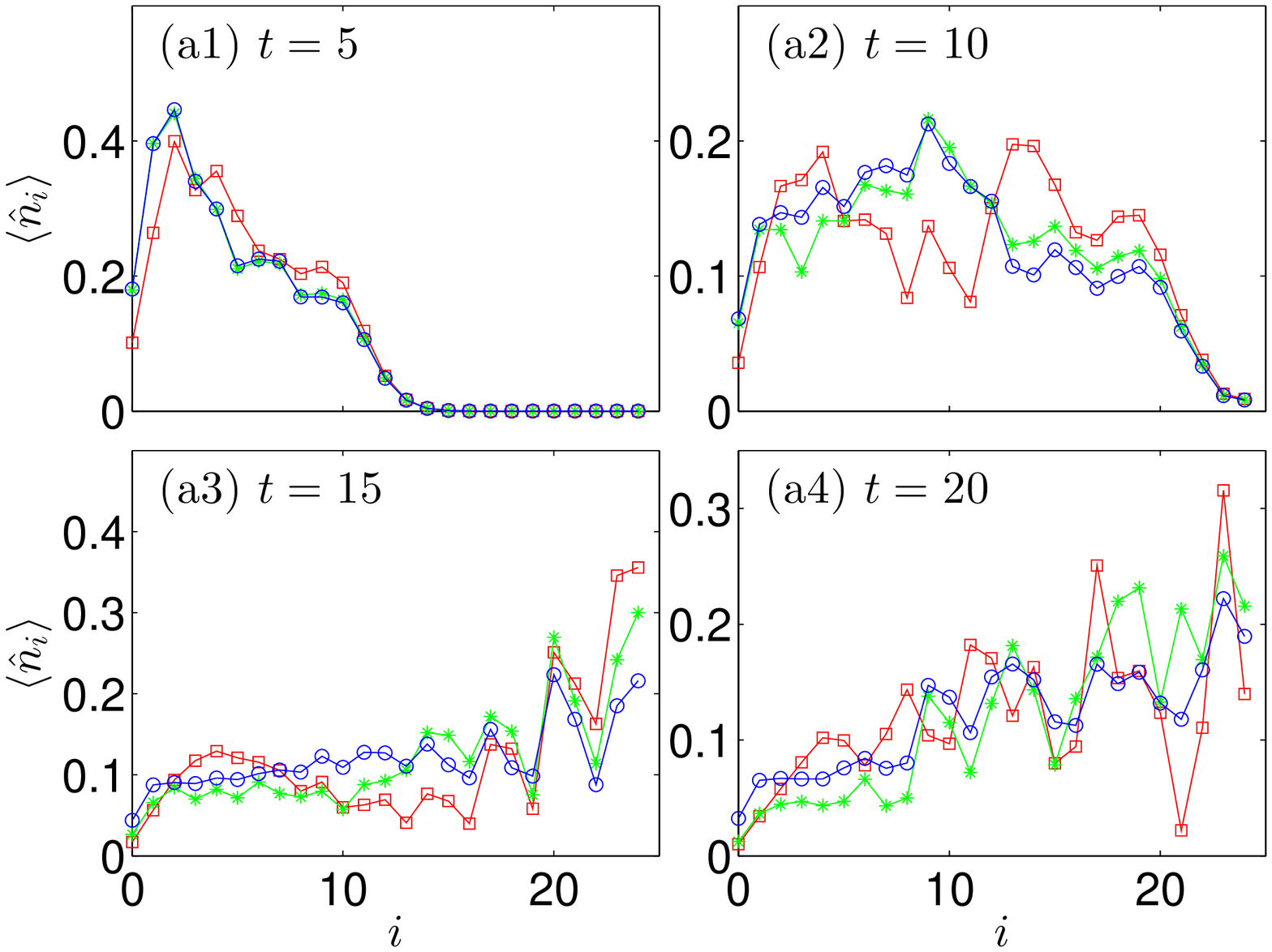}
\includegraphics[width= 0.45 \textwidth ]{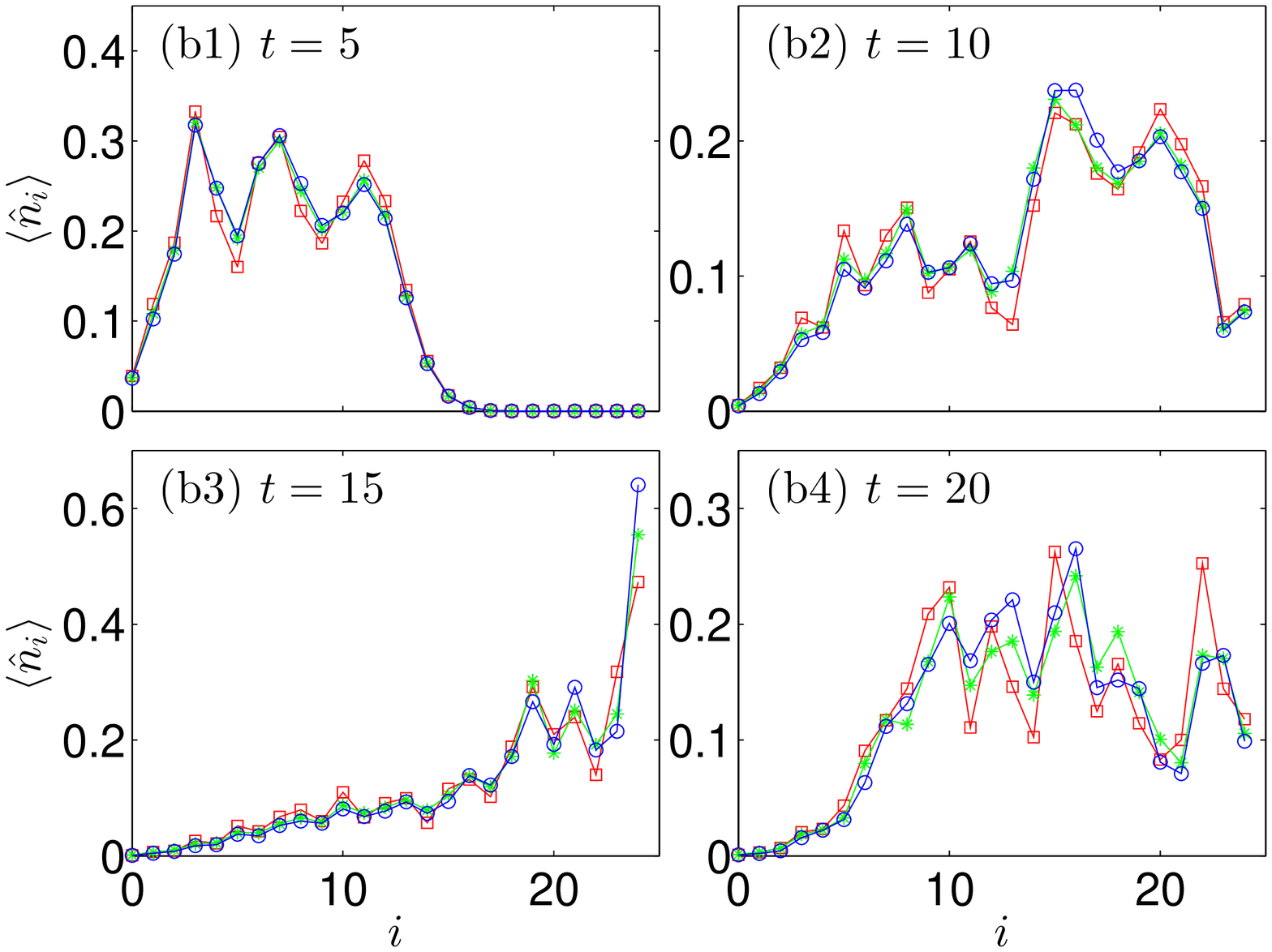}
\caption{(Color online) Time evolution of the density distribution $\langle \hat{n}_i \rangle $ in the two quantum quench scenarios in Fig.~\ref{evoeta}. The left (right) four panels correspond to the left (right) panels in Fig.~\ref{evoeta}. In each panel, the blue line with circles is the result of exact diagonalization and can be considered as exact. The red line with squares is the result of MCTDHF with $M=N=3$ orbitals, while the green line with asterisks is the result of MCTDHF with $M=8$ orbitals. 
\label{den}}
\end{figure*}

The way $\mathcal{I}_\text{max} $ decays has important consequences for the MCTDHF method. In this algorithm, it is assumed that a good approximation of the wave function can be achieved by choosing a limited number of orbitals optimally and adaptively. Now once $\mathcal{I}_\text{max} $, the maximal overlap between a multi-configurational function and the exact wave function, deviates significantly away from unity, the foundation of MCTDHF is unclear and its predictions might be no longer reliable. 

For the two examples in Fig.~\ref{evoeta}, we thus expect the MCTDHF to break down quickly in case (a) while slowly in case (b), which is indeed the case. In Fig.~\ref{den}, the evolution of the density distribution $\langle \hat{n}_i \rangle $ of the system is studied using both exact diagonalization (blue lines with circles) and MCTDHF (red lines with squares, $M=3$; green lines with asterisks, $M=8$). The left four panels [panels (a1)-(a4)] correspond to Fig.~\ref{evoeta}(a), while the right four correspond to Fig.~\ref{evoeta}(b). We see that in case (a) and with $M=3$ orbitals, the prediction of MCTDHF deviates from the exact evolution already significantly at $t=5$ [Fig.~\ref{den}(a1)], and at $t=10$, it misses the main features of the exact distribution [Fig.~\ref{den}(a2)]. Using more orbitals ($M=8$), the result of  MCTDHF improves but beyond $t=15$, it becomes bad again. This is of course not surprising in view of Fig.~\ref{dencomp}(a). There we see that at $t=20$, even the optimal approximate wave functions differ significantly from the exact wave function in density distribution.   

In case (b), on the other hand, even with a single configuration $(M=3)$, MCTDHF can capture the main features of the real density distribution up to $t=20$. With more orbitals ($M=8$), the accuracy improves. This is quite remarkable, since the even with $M=8$, the working dimension of MCTDHF is just 56, which is much smaller than $\mathcal{D}= 2300$, the working dimension of exact diagonalization. Of course, eventually the prediction of MCTDHF will deviate far away from the exact result (not shown here). But this is because of accumulation of error, not because the exact wave function itself cannot be well represented by a multi-configuration approximation. Actually, as we see in Fig.~\ref{dencomp}(b), even at time $t=100$, even in the minimal case ($M=N=3$), the optimal approximate wave function can capture the exact density distribution quantitatively very well. 

\section{Conclusion and Discussion}\label{conclusion}

We have devised an algorithm for constructing the optimal multi-configuration approximation of a multi-fermion wave function. The basic idea is to optimize one orbital with all other orbitals fixed. This procedure can be repeated by sweeping through the orbitals. In this way, the algorithm converges monotonically and thus definitely. Sometimes it can become trapped in local maxima but by running the code several times, the global maximum is hit with high probability. The algorithm can be easily parallelized. We have also provided an upper bound of $\mathcal{I}_\text{max}$ in Eq.~(\ref{myine}). Though not very accurate as an estimate in some circumstances (for the case in Fig.~\ref{evoeta}(a), it overestimates $\mathcal{I}_\text{max}$ by more than a factor of two for $M=3$), it is much easier to calculate. 

We have employed the algorithm to study spinless interacting
  one-dimensional lattice fermions, analyzing whether the ground state
  as well as the time-evolving states can be well approximated by
  multi-configuration Slater determinants. For the ground state, the algorithm reveals the CDW order and
  a bound fermionic pair, in the repulsive and attractive interaction case, respectively. The study of dynamics showed that the performance of MCTDHF depends strongly on whether the time-evolving state can be well approximated by a multi-configuration wave function using only a limited number of orbitals. An important observation is that, in some cases, the multi-configuration approximation improves only slowly as the number of orbitals $M$ increases, i.e., the value of $\mathcal{I}_\text{max} $ increases only slowly with $M$. This implies that it can be inefficient to improve the quality of MCTDHF by increasing the number of orbitals.   

In this paper we have applied the algorithm only to a simple model (one-dimensional and with short-range interaction). In the future, hopefully it will find use in more realistic models in atomic or condensed matter physics. For example, 
the Hartree-Fock approximation is used widely to calculate the ground state of atoms or ions or molecules. But can the ground state be well approximated by a single Slater determinant at all? If so, to what extent? Our algorithm may be used to address these questions. There are actually a lot of systems in atomic and condensed matter physics in which correlation between particles is essential. For example, the negative hydrogen ion \cite{chan, rau01}, and the Laughlin  wave function \cite{laughlin} in the fractional quantum Hall effect, to mention a few of the most famous ones. The latter is usually taken as a paradigm of non-Fermi liquid, with the picture of Fermi sea (a Slater determinant wave function) breaking down completely. In other words, the value of $\mathcal{I}_\text{max}$ must be small for the Laughlin wave function. It would be worthwhile to check this point and study the scaling behaviors.

Unfortunately, the algorithm is applicable only to fermions but not to bosons. 
For bosons, instead of determinants, the basis states of a multi-particle system are permanents. Similar to fermions, for bosons the question is how to choose $M$ orbitals so that the given wave function can be best approximated in the $C_{N+M-1}^N$-dimensional space spanned by the permanents. However, the idea of fixing all orbitals except for one and optimizing it no longer works. 
The reason is simply that for bosons, in a permanent, the same orbital can appear more than once  and one no longer obtains a quadratic maximization problem like in Eq.~(\ref{quad}). Note also that permanents are more difficult to calculate than determinants. While  Gaussian elimination can be used to calculate the determinant in polynomial time, this method cannot be used to calculate the permanent.

Another limitation of the algorithm is of course that the wave function needs to be explicitly available. This means that only systems can be studied that are within the capacity of exact diagonalization methods. An open question is whether one can obtain an estimate (say, an upper or lower bound) of the entanglement in a wave function with only some partial information.

\section*{Acknowledgments}

We are grateful to Alex D. Gottlieb and Mathias Nest for seminal discussions and to Duan-lu Zhou, Michael Sekania, and Patrik Thunstr\"om for their helpful comments. This work was supported in part by the Deutsche Forschungsgemeinschaft through TRR 80.

\appendix
\section{Proofs of the $N$-in-$(N+1)$ theorem}\label{appA}
Though this theorem has been known for a long time \cite{ando}, here we still would like to provide some more proofs. It is always good to have new proofs of old theorems, since different proofs provide different insights from different perspectives.

Suppose $f(x_1, x_2, \ldots, x_N)$ is the wave function of a $N$-fermion system with $N+1$ single-particle orbitals available to each fermion. Let $\{ \phi_1, \phi_2, \ldots, \phi_N\}$ be $N$ orthonormal single-particle orbitals, out of which the Slater determinant having the largest overlap with $f$ can be constructed. This basis can be completed by adding one more orbital $\phi_{N+1}$. Now $f$ can be expanded as 
\begin{eqnarray}\label{fexpan}
f = \sum_{j=1}^{N+1} C_j \phi_1 \wedge \ldots \wedge \tilde{\phi_j} \wedge \ldots \wedge \phi_{N+1}.
\end{eqnarray}
Here $\phi_1 \wedge \ldots \wedge \tilde{\phi_j} \wedge \ldots \wedge \phi_{N+1} $ denotes the Slater determinant constructed out of the $\phi$'s except for $\phi_j$, with the convention that the coefficient of $\phi_1(x_1) \ldots \phi_{j-1}(x_{j-1}) \phi_{j+1} (x_j) \ldots \phi_{N+1}(x_N)$ being positive.

 Now we claim that $C_{j} = 0$ for $j\neq N+1$, and thus the wave function $f$ can be written as a single Slater determinant $\phi_1 \wedge \phi_2 \wedge \ldots \wedge \phi_N$. To prove this, suppose $C_k \neq 0$ for some $1\leq k \leq N$. We then note that the superposition of $C_{N+1} \phi_1 \wedge \phi_2 \wedge \ldots \wedge \phi_N $
and $C_k \phi_1 \wedge  \ldots \wedge \tilde{\phi}_k \wedge \ldots \wedge \phi_{N+1}$ can be rewritten as a single Slater determinant 
\begin{eqnarray}
\sqrt{|C_{N+1}|^2 + |C_k|^2} \phi_1 \wedge  \ldots \wedge {\phi}_k^{mod} \wedge \ldots \wedge \phi_{N}.
\end{eqnarray}
Here $\phi_k^{mod}$ is the modified $k$-th orbital 
\begin{eqnarray}
\phi_k^{mod } = \frac{ C_N \phi_k + (-1)^{N-k} C_k \phi_{N+1}}{\sqrt{|C_{N+1}|^2 + |C_k|^2} },
\end{eqnarray} 
which is a hybrid of $\phi_k$ and $\phi_{N+1}$. We have thus constructed a new Slater determinant with even larger overlap with $f$ than $\phi_1 \wedge \phi_2 \wedge \ldots \wedge \phi_N$. This is in contradiction with the assumption and therefore the claim is proven. 

Another more constructive proof goes like this. Take an arbitrary orthonormal basis $\{ \varphi_1 ,\varphi_2 ,\ldots, \varphi_{N+1} \}$. Similar to (\ref{fexpan}), we have the expansion
\begin{eqnarray}\label{fexpan2}
f = \sum_{j=1}^{N+1} A_j \varphi_1 \wedge \ldots \wedge \tilde{\varphi_j} \wedge \ldots \wedge \varphi_{N+1}.
\end{eqnarray}
Now construct a single-particle orbital 
\begin{eqnarray}\label{hole}
\varphi_{hole} = \sum_{j=1}^{N+1} (-1)^j A_j^* \varphi_j.
\end{eqnarray}
It is straightforward to show that 
\begin{eqnarray}
\int d x_N f(x_1, x_2, \ldots, x_N) \varphi_{hole}^*(x_N) =0.
\end{eqnarray}
It is readily seen that this means that $f$ is the Slater determinant constructed out of the $N$-dimensional subspace complement to $\varphi_{hole}$. The orbital $\varphi_{hole}$ is unoccupied at all and that explains the subscript. 

In the second quantization formalism, the function in (\ref{fexpan2}) is
\begin{eqnarray}
|f \rangle \propto \sum_{j=1}^{N+1} (-1)^{j} A_j c_j c_1^\dagger c_2^\dagger \ldots c_{N+1}^\dagger |0\rangle  .
\end{eqnarray}
Here $|0\rangle$ is the vacuum state and $c_j^\dagger $ ($c_j$) is the creation (annihilation) operator for a fermion in orbital $\varphi_j$. Defining a new creation operator corresponding to the orbital (\ref{hole}),
$c_{hole}^\dagger = \sum_{j=1}^{N+1} (-1)^j A_j^* c_j^\dagger $,
and defining a set of creation operators $\{ \tilde{c}_1^\dagger, \tilde{c}_2^\dagger, \ldots, \tilde{c}_N^\dagger \}$ corresponding to the complement space of $\varphi_{hole}$, we can rewrite $|f\rangle $ as 
\begin{eqnarray}
|f\rangle &\propto & c_{hole} c_1^\dagger c_2^\dagger \ldots c_{N+1}^\dagger |0\rangle  \propto  c_{hole} c_{hole}^\dagger \tilde{c}_1^\dagger \tilde{c}_2^\dagger \ldots \tilde{c}_{N}^\dagger |0\rangle   \nonumber \\
&\propto &  \tilde{c}_1^\dagger \tilde{c}_2^\dagger \ldots \tilde{c}_{N}^\dagger |0\rangle  .
\end{eqnarray}
Therefore, $f$ is a Slater determinant spanned by the complement of $\varphi_{hole}$. Yet another way to understand the ``hole'' is to study the one-particle reduced density matrix. It is easily shown that 
\begin{eqnarray}
\rho_{ji}^{(1)}=\langle f| c_i^\dagger c_j |f\rangle &=& \delta_{ij} -  (-1)^j A_j^* (-1)^{i} A_i.
\end{eqnarray}
It is the identity matrix deflated by one rank, and its kernel is just the hole orbital. 

\section{Closeness in inner product means closeness in density distribution}\label{appB}
Suppose two (normalized to unity) $N$-fermion wave functions $f_1(x_1, \ldots, x_N)$ and $f_2(x_1, \ldots, x_N)$ are close by inner product, i.e.,  
$1-\int dx_1 \ldots dx_N f_1^* f_2 =  \epsilon \ll 1$.
Note that here it is assumed that the inner product is a positive 
real number, as is always achievable by choosing the global phase of $f_1$ or $f_2$. A direct consequence is
$1- \int dx_1 \ldots dx_N |f_1||f_2| \leq \epsilon $.
The corresponding density distribution is given by ($i=1$, 2)
\begin{eqnarray}
n_i (x_1) = N \int dx_2 \ldots d x_N |f_i(x_1, x_2, \ldots, x_N)|^2.
\end{eqnarray}
We have for the $L_1$ distance between $n_1$ and $n_2$,
\begin{eqnarray}
\delta_1 & \equiv & \frac{1}{N}\int d x_1  |n_1(x_1) - n_2 (x_1)| \nonumber \\
 &\leq &  \int dx_1 \ldots d x_N \left||f_1| - |f_2| \right|  \left(|f_1|+|f_2| \right ) \nonumber \\ 
  &\leq & \sqrt{\int \prod_{j=1}^N dx_j (|f_1| - |f_2|)^2\int \prod_{j=1}^N dx_j (|f_1| + |f_2|)^2}  \nonumber \\
 &\leq & \sqrt{8 \epsilon }.
\end{eqnarray}
Therefore, closeness in inner product means closeness in density distribution. The reverse is obviously not true.


\begin{thebibliography}{99}

\bibitem{slater}
J. C. Slater, Phys. Rev. \textbf{34}, 1293 (1929).

\bibitem{noquan}
Since an unambiguous definition of correlation or entanglement in a multipartite system is still missing, and it is the case especially for identical particles, in this paper we talk about correlation or entanglement qualitatively or intuitively only. 

\bibitem{yukalov}
A.~J. Coleman and V.~I. Yukalov, \textit{Reduced Density Matrices: Coulson's Challenge} (Springer, Heidelberg, 2000).

\bibitem{ando}
T. Ando, Rev. Mod. Phys. \textbf{35}, 690 (1963).

\bibitem{coleman}
A.~J. Coleman, Rev. Mod. Phys. \textbf{35}, 668 (1963).

\bibitem{alex}
A.~D. Gottlieb and N.~J. Mauser, Phys. Rev. Lett. \textbf{95}, 123003 (2005).

\bibitem{alex2}
A.~D. Gottlieb and N.~J. Mauser, Int. J. Quantum Information \textbf{5}, 815 (2007).

\bibitem{buchleitner}
M. C Tichy, F. Minter, and A. Buchleitner, J. Phys. B: At. Mol. Opt. Phys. \textbf{44}, 192001 (2011).

\bibitem{vollhardt}
K. Byczuk, J. Kune\v{s}, W. Hofstetter, and D. Vollhardt, Phys. Rev. Lett. \textbf{108}, 087004 (2012); \textit{ibid.} \textbf{108}, 189902 (2012).

\bibitem{eriksson}
P. Thunstr\"om, I. Di Marco, and O. Eriksson, Phys. Rev. Lett. \textbf{109}, 186401 (2012).

\bibitem{zanghellini}
J. Zanghellini, M. Kitzler, C. Fabian, T. Brabec, and A. Scrinzi, Laser Phys. \textbf{13}, 1064 (2003).

\bibitem{caillat}
J. Caillat, J. Zanghellini, M. Kitzler, O. Koch, W. Kreuzer, and A. Scrinzi, Phys. Rev. A \textbf{71}, 012712 (2005).

\bibitem{kato}
T. Kato and H. Kono, Chem. Phys. Lett. \textbf{392}, 533 (2004).

\bibitem{nest}
M. Nest, T. Klamroth, and P. Saalfrank, J. Chem. Phys. \textbf{122}, 124102 (2005).

\bibitem{alon}
O.~E. Alon, A.~I. Streltsov, and L.~S. Cederbaum, J. Chem. Phys. \textbf{127}, 154103 (2007).

\bibitem{sakmann}
A.~U.~J. Lode, K. Sakmann, O.~E. Alon, L.~S. Cederbaum, and A.~I. Streltsov, Phys. Rev. A \textbf{86}, 063606 (2012).

\bibitem{uppsala}
A recursive algorithm was proposed in \cite{eriksson}, but only for the single configuration case, i.e., $M=N$.

\bibitem{alexnote}
A.~D. Gottlieb (private communication).

\bibitem{brueckner1}
K.~A. Brueckner and C.~A. Levinson, Phys. Rev. \textbf{97}, 1344 (1955).

\bibitem{brueckner2}
K.~A. Brueckner and W. Wada, Phys. Rev. \textbf{103}, 1008 (1956).

\bibitem{nesbet}
R.~K. Nesbet, Phys. Rev. \textbf{109}, 1632 (1958).

\bibitem{smith}
S. Larsson and V.~H. Smith, Jr., Phys. Rev. \textbf{178}, 137 (1969).

\bibitem{lowdin}
P.-O. L\"owdin and H. Shull, Phys. Rev. \textbf{101}, 1730 (1956).

\bibitem{rmp57}
H. Everett, Rev. Mod. Phys. \textbf{29}, 454 (1957). 

\bibitem{eberly}
R. Grobe, K. Rzazewski, and J.~H. Eberly, J. Phys. B: At. Mol. Opt. Phys. \textbf{27}, L503 (1994).

\bibitem{matrix}
R. Bhatia, \textsl{Matrix Analysis} (Springer-Verlag, New York, 1997).

\bibitem{golub}
G.~H. Golub and C.~F.~Van Loan, \textsl{Matrix Computations} (Johns Hopkins University Press, 3rd edition, 1996).

\bibitem{NO}
A natural idea is to use the natural orbitals as the initial guess of the $\phi$'s. For $N=2$, they are simply the ultimate answer. However, for $N\geq 3$, numerical experiments showed that they do not offer any definite advantage to randomly chosen initial guesses. 

\bibitem{circle}
R.~A. Horn and C.~R. Johnson, \textsl{Matrix Analysis} (Cambridge University Press, 1999).

\bibitem{scaling}
The $1/L$ scaling behavior is not visible in Fig.~\ref{gsnegative}(a) because $U=-2.05$ is too close to $U_c =-2$. However, it is verified for those $U$ farther away from $U_c$.

\bibitem{ejp}
J.~M.~Zhang and R.~X.~Dong, Eur.~J.~Phys. \textbf{31}, 591 (2010).

\bibitem{right}
This proves that the algorithm works correctly and the global maximum has been found. 

\bibitem{chan}
S. Chandrasekhar, Astrophys. J. \textbf{100}, 176 (1944).

\bibitem{rau01}
A. R. P. Rau, Am. J. Phys. \textbf{80}, 406 (2012); A. R. P. Rau, J. Astrophys. Astr. \textbf{17}, 113 (1996).

\bibitem{laughlin}
R. B. Laughlin, Phys. Rev. Lett. \textbf{50}, 1395 (1983).


\end{thebibliography}
\end{document}